\documentclass{sigplanconf}
\usepackage{amsmath}
\usepackage[british]{babel}
\usepackage[lighttt]{lmodern}
\usepackage{dsfont}
\usepackage{textcomp}
\usepackage{cancel}
\usepackage[inline]{enumitem}
\usepackage{alltt}
\usepackage{tikz}

\usepackage{amssymb}
\usepackage{amsmath}

\newtheorem{theorem}{Theorem}

\newcommand{\textapprox}{\raise.17ex\hbox{$\scriptstyle\mathtt{\sim}$}}

\usepackage{doc}
\usepackage{makeidx}

\usepackage{color}
\usepackage{fancyvrb}
\usepackage{proof}
\usepackage{url}

\usepackage{listings}
\lstdefinelanguage{tptp}{
  basicstyle={\ttfamily},
  columns=flexible,
  keywords={tff, type, axiom, hypothesis, conjecture},
  sensitive=true,
  captionpos=b,
}
\lstdefinelanguage{cpp}{
  basicstyle={\ttfamily},
  columns=flexible,
  keywords={public, static, void, do, for, while, int, bool, if, else, break},
  sensitive=true,
  captionpos=b
}

\newcommand{\reserved}[1]{\textbf{\underline{#1}}} 
\newcommand{\ass}{\texttt{:=}}     
\newcommand{\inc}{~~~~\= \+ \kill}    
\newcommand{\dec}{\- \kill}         
\newcommand{\semicol}{;}                  

\newcommand{\IF}{\reserved{if}}

\newcommand{\DO}{\reserved{do}}
\newcommand{\OD}{\reserved{end~do}}
\newcommand{\THEN}{\reserved{then}}
\newcommand{\ELSE}{\reserved{else}}
\newcommand{\WHILE}{\reserved{while}}

\newcommand{\ASS}{\texttt{ := }}

\newcommand{\nofoolVampire}{Vampire$\,\star$}

\newcommand{\ite}[3]{\mathtt{if}\;{#1}\;\mathtt{then}\;{#2}\;\mathtt{else}\;{#3}}

\newcommand{\ITE}{\texttt{if-then-else}}
\newcommand{\binding}[2]{{#1}\,=\,{#2}}
\newcommand{\letin}[3]{\mathtt{let}\;\binding{#1}{#2}\;\mathtt{in}\;{#3}}
\newcommand{\letinpar}[5]{\mathtt{let}\;\binding{#1}{#2};\;\binding{#3}{#4}\;\mathtt{in}\;{#5}}
\newcommand{\LETIN}{\texttt{let-in}}

\newcommand{\true}{\mathit{true}}
\newcommand{\false}{\mathit{false}}
\newcommand{\bool}{\mathit{bool}}
\newcommand{\fool}{{FOOL}}
\newcommand{\foolp}{{FOOL+}}

\renewcommand{\implies}{\Rightarrow}
\newcommand{\liff}{\Leftrightarrow}

\newcommand{\eql}{\doteq}
\newcommand{\neql}{\not\doteq}

\newcommand{\tptpo}{\lstinline'$o'} 
\newcommand{\dbool}{\lstinline'$bool'} 
\newcommand{\dtrue}{\lstinline'$true'} 
\newcommand{\dfalse}{\lstinline'$false'} 
\newcommand{\ddtrue}{\lstinline'$$true'}
\newcommand{\ddfalse}{\lstinline'$$false'}

\newcommand{\dite}{\lstinline'$ite'} 
\newcommand{\ditet}{\lstinline'$ite_t'} 
\newcommand{\ditef}{\lstinline'$ite_f'} 

\newcommand{\dlet}{\lstinline'$let'} 
\newcommand{\dlettt}{\lstinline'$let_tt'} 
\newcommand{\dlettf}{\lstinline'$let_tf'} 
\newcommand{\dletft}{\lstinline'$let_ft'} 
\newcommand{\dletff}{\lstinline'$let_ff'} 

\newcommand{\dint}{\lstinline'$int'} 
\newcommand{\dgreatereq}{\lstinline'$greatereq'} 
\newcommand{\dsum}{\lstinline'$sum'} 

\newcommand{\arrayt}{\mathit{array}}
\newcommand{\select}[2]{\mathit{select}({#1},{#2})}
\newcommand{\selectf}{\mathit{select}}
\newcommand{\store}[3]{\mathit{store}({#1},{#2},{#3})}
\newcommand{\storef}{\mathit{store}}

\newcommand{\darray}[2]{\darraySymb\lstinline'('{#1}\lstinline','{#2}\lstinline')'}
\newcommand{\darraySymb}{\lstinline'$array'} 
\newcommand{\dselect}{\lstinline'$select'} 
\newcommand{\dstore}{\lstinline'$store'} 
\newcommand{\darrayone}{\lstinline'$array1'} 
\newcommand{\dselectone}{\lstinline'$select1'} 
\newcommand{\dstoreone}{\lstinline'$store1'} 
\newcommand{\darraytwo}{\lstinline'$array2'} 
\newcommand{\dselecttwo}{\lstinline'$select2'} 
\newcommand{\dstoretwo}{\lstinline'$store2'} 

\newcommand{\Z}{\mathds{Z}}

\begin{document}

\setlength{\pdfpageheight}{\paperheight}
\setlength{\pdfpagewidth}{\paperwidth}

\conferenceinfo{CPP '16}{January 18--19, 2016, Saint Petersburg, Florida, USA}
\copyrightyear{2016}
\copyrightdata{978-1-nnnn-nnnn-n/yy/mm}
\doi{nnnnnnn.nnnnnnn}

\newcommand{\EK}[1]{{\color{red} EK: {#1}}}
\newcommand{\GR}[1]{{\color{green} GR: {#1}}}
\newcommand{\LK}[1]{{\color{blue} LK: {#1}}}
\newcommand{\AV}[1]{{\color{blue} AV: {#1}}}

%
\title{The Vampire and the FOOL}



%
\authorinfo{Evgenii Kotelnikov}{Chalmers University of Technology, Gothenburg, Sweden}{evgenyk@chalmers.se}
\authorinfo{Laura Kov\'{a}cs}{Chalmers University of Technology, Gothenburg, Sweden}{laura.kovacs@chalmers.se}
\authorinfo{Giles Reger}{The University of Manchester, Manchester, UK}{giles.reger@manchester.ac.uk}
\authorinfo{Andrei Voronkov}{The University of Manchester; Chalmers University of Technology; EasyChair}{andrei@voronkov.com}


\clearpage

\maketitle

\begin{abstract}
  This paper presents new features recently implemented in the theorem prover Vampire, namely support for first-order logic with a first class boolean sort (FOOL) and polymorphic arrays. In addition to having a first class boolean sort, FOOL also contains \ITE\ and \LETIN\ expressions. We argue that presented extensions facilitate reasoning-based program analysis, both by increasing the expressivity of first-order reasoners and by gains in efficiency.
\end{abstract}



%
%


\section{Introduction}
\label{sect:introduction}

Automated program analysis and verification requires discovering and proving program properties. These program properties are checked using various tools, including theorem provers. The translation of program properties into formulas accepted by a theorem prover is not straightforward because of a mismatch between the semantics of the programming language constructs and that of the input language of the theorem prover. If program properties are not directly expressible in the input language, one should implement a translation of such program properties to the language. Such translations can be very complex and thus error prone.

The performance of a theorem prover on the result of a translation crucially depends on whether the translation introduces formulas potentially making the prover inefficient. Theorem provers, especially first-order ones, are known to be very fragile with respect to the input. Expressing program properties in the ``right'' format therefore requires solid knowledge about how theorem provers work and are implemented~--- something that a user of a verification tool might not have. Moreover, it can be hard to efficiently reason about certain classes of program properties, unless special inference rules and heuristics are added to the theorem prover. For example, \cite{ATVA14} shows a considerable gain in performance on proving properties of data collections by using a specially designed extensionality resolution rule.

If a theorem prover natively supports expressions that mirror the semantics of programming language constructs, we solve both above mentioned problems. First, the users do not have to design translations of such constructs. Second, the users do not have to possess a deep knowledge of how the theorem prover works~--- the efficiency becomes the responsibility of the prover itself.

In this work we present new features recently developed and implemented in the theorem prover Vampire~\cite{cav13} to natively support mirroring programming language constructs in its input language. They include (i) FOOL~\cite{FOOL}, that is the extension of first-order logic by a first-class boolean sort, \ITE\ and \LETIN\ expressions, and (ii)  polymorphic arrays.

This paper is structured as follows. Section~\ref{sect:fool} presents how FOOL is implemented in Vampire and focuses on new extensions to the TPTP input language~\cite{TPTP} of first-order provers. Section~\ref{sect:fool}  extends the TPTP language of monomorphic many-sorted first-order formulas, called TFF0~\cite{tff0}, and allows users to treat the built-in boolean sort \tptpo\ as a first class sort. Moreover, it introduces expressions \dite\ and \dlet, which unify various TPTP \ITE\ and \LETIN\ expressions.

Section~\ref{sect:arrays} presents a formalisation of a polymorphic theory of arrays in TPTP and its implementation in Vampire.
It extends TPTP  with features of the TFF1 language~\cite{tff1} of rank-1 polymorphic many-sorted first-order formulas,  namely, sort arguments for the built-in array sort constructor \darraySymb. Sort variables however are not supported.

We argue that these extensions make the translation of properties of some programs to TPTP easier. To support this claim, in Section~\ref{sect:example} we discuss representation of various programming and other constructs in the extended TPTP language. We also give a linear translation of  the next state relation for any program with assignments, \ITE, and sequential composition.

Experiments with theorem proving with FOOL formulas are described in Section~\ref{sect:experiments}. In particular, we show that the implementation of a new inference rule, called FOOL paramodulation, improves performance of theorem provers using superposition calculus.

Finally, Section~\ref{sect:related} discusses related work and Section~\ref{sect:future} outlines future work.


\noindent\paragraph{Summary of the main results.}
\begin{itemize}
\item We describe an implementation of first-order logic with a first-class boolean sort. This bridges the gap between input languages for theorem provers and logics and tools used in program analysis. We believe it is a first ever implementation of first-class boolean sorts in superposition theorem provers.

\item We extend and simplify the TPTP language~\cite{TPTP}, by providing more powerful and more uniform representations of \ITE\ and \LETIN\ expressions. To the best of our knowledge, Vampire is the only superposition theorem prover implementing these constructs.

\item We formalise and describe an implementation in Vampire of a polymorphic theory of arrays. Again, we believe that Vampire is the only superposition theorem prover implementing this theory.

\item We give a simple extension of FOOL, allowing to express the next state relation of a program as a boolean formula which is linear in the size of the program. This  boolean formula captures the exact semantics of the program and can be used by a first-order theorem prover. We are not aware of any other work on extending theorem provers with support for representing fragments of imperative programs.

\item We demonstrate usability and high performance of our implementation on two collections of examples, coming from the higher-order part of the TPTP library and from the Isabelle interactive theorem prover~\cite{Isabelle}. Our experimental results show that Vampire outperforms systems which could previously be used to solve such problems:
higher-order theorem provers and satisfiability modulo theory (SMT)  solvers.
\end{itemize}

The paper focuses on new, practical features extending first-order theorem provers for making them better suited for applications of reasoning in various theories, program analysis and verification. While the paper describes implementation details and challenges in the Vampire theorem prover, the described features and their implementation can be carried out in any other first-order prover.

Summarising, we believe that our paper advances the state-of-the-art in formal certification of programs and proofs. With the use of FOOL and polymorphic arrays, we bring first-order theorem proving closer to program logics and make first-order theorem proving better suited for program analysis and verification. We also believe that an implementation of FOOL advances automation of mathematics, making many problems using the boolean type directly understood by a first-order theorem prover, while they previously were treated as higher-order problems.

\section{First Class Boolean Sort}
\label{sect:fool}

Our recent work~\cite{FOOL} presented a modification of many-sorted first-order logic that contains a boolean sort with a fixed interpretation and treats terms of the boolean sort as formulas. We called this logic FOOL, standing for first-order logic (FOL) + boolean sort. FOOL extends FOL by (1) treating boolean terms as formulas; (2) \ITE\ expressions; and (3) \LETIN\ expressions. There is a model-preserving transformation of FOOL formulas to FOL formulas, hence an implementation of this transformation makes it possible to prove FOOL formulas using a first-order theorem prover.

The language of FOOL is, essentially, a superset of the core language of SMT-LIB~2~\cite{SMT-LIB}, the library of problems for SMT solvers. The difference between FOOL and the core language is that the former has richer \LETIN\ expressions, which support local definitions of functions symbols of arbitrary arity, while the latter only supports local binding of variables.

FOOL can be regarded as the smallest superset of the SMT-LIB~2 Core language and TFF0. An implementation of a translation of FOOL to FOL thus also makes it possible to translate SMT-LIB problems to TPTP. Consider, for example, the following tautology, written in the SMT-LIB syntax: \verb'(exists ((x Bool)) x)'. It quantifies over boolean variables and uses a boolean variable as a formula. Neither is allowed in the standard TPTP language, but can be directly expressed in an extended TPTP that represents FOOL.

The rest of this section presents features of FOOL not included in FOL, explains how they are implemented in Vampire and how they can be represented in an extended TPTP syntax understood by Vampire.

\subsection{Proving with the Boolean Sort}

Vampire supports many-sorted predicate logic and the TFF0 syntax for this logic. In many-sorted predicate logic all sorts are uninterpreted, while the boolean sort should be interpreted as a two-element set. There are several ways to support the boolean sort in a first-order theorem prover, for example, one can axiomatise it by adding two constants $\true$ and $\false$ of this sort and two axioms: $(\forall x:\bool)(x \eql \true \lor x \eql \false)$ and $\true \neql \false$. However, as we discuss in \cite{FOOL}, using this axiomatisation in a superposition theorem prover may result in performance problems caused by self-paramodulation of $x \eql \true \lor x \eql \false$.

To overcome this problem, in \cite{FOOL} we proposed the following modification of the superposition calculus.
\begin{enumerate}
  \item Use a special simplification ordering that makes the constants $\true$ and $\false$ smallest terms of the sort $\bool$ and also makes $\true$ greater than $\false$.

\item Add the axiom $\true \neql \false$.

\item Add a special inference rule, called \emph{FOOL paramodulation}, of the form
  \[
    \infer[,]{C[\mathtt{\true}] \lor s \eql \mathtt{\false}}{C[s]}
  \]
where
\begin{enumerate}
\item $s$ is a term of the sort $\bool$ other than $\true$ and $\false$;
\item $s$ is not a variable;
\end{enumerate}
\end{enumerate}

Both ways of dealing with the boolean sort are supported in Vampire. They are controlled by the option \verb|--fool_paramodulation|, which can be set to \verb|on| or \verb|off|. The default value is \verb|on|, which enables the modification.

Vampire uses the TFF0 subset of the TPTP syntax, which does not fully support FOOL. To write FOOL formulas in the input, one uses the standard TPTP notation: \tptpo\ for the boolean sort, \dtrue\ for $\true$ and \dfalse\ for $\false$. There are, however, two ways to output the boolean sort and the constants. One way will use the same notation as in the input and is the default, which is sufficient for most applications. The other way can be activated by the option \verb'--show_fool on', it will
\begin{enumerate}
  \item denote as \dbool\ every occurrence of $\bool$ as a sort of a variable or an argument (to a function or a predicate symbol);
  \item denote as \ddtrue\ every occurrence of $\true$ as an argument; and
  \item denote as \ddfalse\ every occurrence of $\false$ as an argument.
\end{enumerate}
Note that an occurrence of any of the symbols \dbool, \ddtrue\ or \ddfalse\ anywhere in an input problem is not recognised as syntactically correct by Vampire.

Setting \verb'--show_fool' to \verb'on' might be necessary if Vampire is used as a front-end to other reasoning tools. For example, one can use Vampire not only for proving, but also for preprocessing the input problem or converting it to clausal normal form. To do so, one uses the options \verb|--mode preprocess| and \verb|--mode clausify|, respectively. The output of Vampire can then be passed to other theorem provers, that either only deal with clauses or do not have sophisticated preprocessing. Setting \verb'--show_fool' to \verb'on' appends a definition of a sort denoted by \dbool\ and constants denoted by \ddtrue\ and \ddfalse\ of this sort to the output. That way the output will always contain syntactically correct TFF0 formulas, which might not be true if the option is set to \verb'off' (the default value).

Every formula of the standard FOL is syntactically a FOOL formula and has the same models. Vampire does not reason in FOOL natively, but rather translates the input FOOL formulas into FOL formulas in a way that preserves models. This is done at the first stage of preprocessing of the input problem.

Vampire implements the translation of FOOL formulas to FOL given in~\cite{FOOL}. It involves replacing parts of the problem that are not syntactically correct in the standard FOL by applications of fresh function and predicate symbols. The set of assumptions is then extended by formulas that define these symbols. Individual steps of the translation are displayed when the \verb'--show_preprocessing' options is set to \verb'on'.

In the next subsections we present the features of FOOL that are not present in FOL together with their syntax in the extended TFF0 and their implementation in Vampire.

\subsection{Quantifiers over the Boolean Sort}

FOOL allows quantification over $\bool$ and usage of boolean variables as formulas. For example, the formula $(\forall x:\bool)(x \lor \neg x)$ is a syntactically correct tautology in FOOL. It is not however syntactically correct in the standard FOL where variables can only occur as arguments.

Vampire translates boolean variables to FOL in the following way. First, every formula of the form $x \liff y$, where $x$ and $y$ are boolean variables, is replaced by $x \eql y$. Then, every occurrence of a boolean variable $x$ anywhere other than in an argument is replaced by $x \eql \true$. For example, the tautology $(\forall x:\bool)(x \lor \neg x)$ will be converted to the FOL formula $(\forall x:\bool)(x \eql \true \lor x \neql \true)$ during preprocessing.

Note that it is possible to directly express quantified boolean formulas (QBF) in FOOL, and use Vampire to reason about them.

TFF0 does not support quantification over booleans. Vampire supports an extended version of TFF0 where the sort symbol \tptpo\ is allowed to occur as the sort of a quantifier and boolean variables are allowed to occur as formulas. The formula $(\forall x:\bool)(x \lor \neg x)$ can be expressed in this syntax as \lstinline'![X:$o]: (X | ~X)'. 

\subsection{Functions and Predicates with Boolean Arguments}

Functions and predicates in FOOL are allowed to take booleans as arguments. For example, one can define the logical implication as a binary function $\mathit{impl}$ of the type $\bool \times \bool \to \bool$ using the following axiom:
\[
  (\forall x: \bool)(\forall y: \bool)(\mathit{impl}(x, y) \liff \neg x \lor y).
\]

Since Vampire supports many-sorted logic, this feature requires no additional implementation, apart from changes in the parser.

In TFF0, functions and predicates cannot have arguments of the sort \tptpo. In the version of TFF0, supported by Vampire, this restriction is removed. Thus, the definition of $\mathit{impl}$ can be expressed in the following way.
\begin{lstlisting}
tff(impl, type, ($o * $o) > $o).
tff(impl_definition, axiom,
    ![X:$o, Y:$o]: (impl(X,Y) <=> (~X | Y))).
\end{lstlisting}

\subsection{Formulas as Arguments}

Unlike the standard FOL, FOOL does not make a distinction between formulas and boolean terms. It means that a function or a predicate can take a formula as a boolean argument, and formulas can be used as arguments to equality between booleans. For example, with the definition of $\mathit{impl}$, given earlier, we can express in FOOL that
$P$ is a graph of a (partial) function of the type $\sigma \to \tau$ as follows:
\begin{equation}\label{eq:bool-arg-example}
  (\forall x:\sigma)(\forall y:\tau)(\forall z:\tau)\mathit{impl}(P(x,y) \land P(x,z), y \eql z).
\end{equation}

Note that the definition of $\mathit{impl}$ could as well use equality instead of equivalence.

In order to support formulas occurring as arguments, Vampire does the following. First, every expression of the form $\varphi \eql \psi$ is replaced by $\varphi \liff \psi$. Then, for each formula $\psi$ occurring as an argument the following translation is applied. If $\psi$ is a nullary predicate $\top$ or $\bot$, it is replaced by $\true$ or $\false$, respectively. If $\psi$ is a boolean variable, it is left as is. Otherwise, the translation is done in several steps. Let $x_1,\ldots,x_n$ be all free variables of $\psi$ and $\sigma_1,\ldots,\sigma_n$ be their sorts. Then Vampire
\begin{enumerate}
  \item introduces a fresh function symbol $g$ of the type $$\sigma_1 \times \ldots \times \sigma_n \to \bool;$$
  \item adds the definition $$(\forall x_1:\sigma_1)\ldots(\forall x_n:\sigma_n)(\psi \liff g(x_1,\ldots,x_n) \eql \true)$$ to its set of assumptions;
  \item replaces $\psi$ by $g(x_1,\ldots,x_n)$.
\end{enumerate}

For example, after this translation has been applied for both arguments of $\mathit{impl}$, \eqref{eq:bool-arg-example} becomes $$(\forall x:\sigma)(\forall y:\sigma)(\forall z:\sigma)\mathit{impl}(g_1(x, y, z), g_2(y, z)),$$ where $g_1$ and $g_2$ are fresh function symbol of the types $\sigma \times \tau \times \tau \to \bool$ and $\tau \times \tau \to \bool$, respectively, defined by the following formulas:
\begin{enumerate}
  \item $(\forall x:\sigma)(\forall y:\tau)(\forall z:\tau)(P(x,y) \land P(x,z) \liff g_1(x,y,z) \eql \true)$;
  \item $(\forall y:\tau)(\forall z:\tau)(y \eql z \liff g_2(y,z) \eql \true)$.
\end{enumerate}

TFF0 does not allow formulas to occur as arguments. The extended version of TFF0, supported by Vampire, removes this restriction for arguments of the boolean sort. Formula~\eqref{eq:bool-arg-example} can be expressed in this syntax as follows:
\begin{lstlisting}
![X:s, Y:t, Z:t]: impl(p(X,Y) & p(X,Z), Y = Z)
\end{lstlisting}

For a more interesting example, consider the following logical puzzle taken from the TPTP problem \mbox{PUZ081}:
\begin{quote}
  A very special island is inhabited only by knights and knaves. Knights always tell the truth, and knaves always lie. You meet two inhabitants: Zoey and Mel. Zoey tells you that Mel is a knave. Mel says, `Neither Zoey nor I are knaves'. Who is a knight and who is a knave?
\end{quote}

\newcommand{\knight}{\mathit{Knight}}
\newcommand{\knave}{\mathit{Knave}}
\newcommand{\says}{\mathit{Says}}
\newcommand{\statement}{\mathit{statement}}
\newcommand{\person}{\mathit{person}}
\newcommand{\zoye}{\mathit{zoye}}
\newcommand{\mel}{\mathit{mel}}
To solve the puzzle, one can formalise it as a problem in FOOL and give a corresponding extended TFF0 representation to Vampire. Let $\zoye$ and $\mel$ be terms of a fixed sort $\person$ that represent Zoye and Mel, respectively. Let $\says$ be a predicate that takes a term of the sort $\person$ and a boolean term. We will write $\says(p, s)$ to denote that a person $p$ made a logical statement $s$. Let $\knight$ and $\knave$ be predicates that take a term of the sort $\person$. We will write $\knight(p)$ or $\knave(p)$ to denote that a person $p$ is a knight or a knave, respectively. We will express the fact that knights only tell the truth and knaves only lie by axioms $(\forall p:\person)(\forall s:\bool)(\knight(p) \land \says(p, s) \implies s)$ and $(\forall p:\person)(\forall s:\bool)(\knave(p) \land \says(p, s) \implies \neg s)$, respectively. We will express the fact that every person is either a knight or a knave by the axiom $(\forall p:\person)(\knight(p) \oplus \knave(p))$, where $\oplus$ is the ``exclusive or'' connective. Finally, we will express the statements that Zoye and Mel make in the puzzle by axioms $\says(\zoye, \knave(\mel))$ and $\says(\mel, \neg\knave(\zoye) \land \neg\knave(\mel))$, respectively.

The axioms and definitions, given above, can be written in the extended TFF0 syntax in the following way.
\begin{lstlisting}
tff(person, type, person: $tType).
tff(says, type, says: (person * $o) > $o).

tff(knight, type, knight: person > $o).
tff(knights_always_tell_truth, axiom,
    ![P:person, S:$o]:
      (knight(P) & says(P, S) => S)).

tff(knave, type, knave: person > $o).
tff(knaves_always_lie, axiom,
    ![P:person, S:$o]:
      (knave(P) & says(P, S) => ~S)).

tff(very_special_island, axiom,
    ![P:person]: (knight(P) <~> knave(P))).

tff(zoey, type, zoey: person).
tff(mel,  type, mel:  person).

tff(zoye_says, hypothesis,
    says(zoey, knave(mel))).

tff(mel_says, hypothesis,
    says(mel, ~knave(zoey) & ~knave(mel))).
\end{lstlisting}

Vampire accepts this code, finds that the problem is satisfiable and outputs the saturated set of clauses. There one can see that Zoey is a knight and Mel is a knave. Note that the existing formalisations of this puzzle in TPTP (files \verb'PUZ081^1.p', \verb'PUZ081^2.p' and \verb'PUZ081^3.p') employ the language of higher-order logic (THF)~\cite{THF}. However, as we have just shown, one does not need to resort to reasoning in higher-order logic for this problem, and can enjoy the efficiency of reasoning in first-order logic.

%
%
%
%
%
%
%
%

This example makes one think about representing sentences in various epistemic or first-order modal logics in FOOL.

\subsection{\ITE}

FOOL contains expressions of the form $\ite{\psi}{s}{t}$, where $\psi$ is a boolean term, and $s$ and $t$ are terms of the same sort. The semantics of such expressions mirrors the semantics of conditional expressions in programming languages.

\ITE\ expressions are convenient for expressing formulas coming from program analysis and interactive theorem provers. For example, consider the $\mathit{max}$ function of the type $\Z \times \Z \to \Z$ that returns the maximum of its arguments. Its definition can be expressed in FOOL as
\begin{equation}\label{eq:ite-t-example}
  (\forall x:\Z)(\forall y:\Z)(\mathit{max}(x, y) \eql \ite{x \geq y}{x}{y}).
\end{equation}

To handle such expressions, Vampire translates them to FOL. This translation is done in several steps. Let $x_1,\ldots,x_n$ be all free variables of $\psi$, $s$ and $t$, and $\sigma_1,\ldots,\sigma_n$ be their sorts. Let $\tau$ be the sort of both $s$ and $t$. The steps of translation depend on whether $\tau$ is $\bool$ or a different sort. If $\tau$ is not $\bool$, Vampire
\begin{enumerate}
  \item introduces a fresh function symbol $g$ of the type $$\sigma_1 \times \ldots \times \sigma_n \to \tau;$$
  \item adds the definitions $$(\forall x_1:\sigma_1)\ldots(\forall x_n:\sigma_n) (\psi \implies g(x_1,\ldots,x_n) \eql s)$$ and $$(\forall x_1:\sigma_1)\ldots(\forall x_n:\sigma_n) (\neg\psi \implies g(x_1,\ldots,x_n) \eql t)$$ to its set of assumptions;
  \item replaces $\ite{\psi}{s}{t}$ by $g(x_1,\ldots,x_n)$.
\end{enumerate}

If $\tau$ is $\bool$, the following is different in the steps of translation:
\begin{enumerate}
  \item a fresh predicate symbol $g$ of the type $\sigma_1 \times \ldots \times \sigma_n$ is introduced instead; and
  \item the added definitions use equivalence instead of equality.
\end{enumerate}
\noindent
For example, after this translation \eqref{eq:ite-t-example} becomes $$(\forall x:\Z)(\forall y:\Z)(\mathit{max}(x, y) \eql g(x, y)),$$ where $g$ is a fresh function symbol of the type $\Z \times \Z \to \Z$ defined by the following formulas:
\begin{enumerate}
  \item $(\forall x:\Z)(\forall y:\Z)(x \geq y \implies g(x, y) \eql x)$;
  \item $(\forall x:\Z)(\forall y:\Z)(x \not\geq y \implies g(x, y) \eql y).$
\end{enumerate}

TPTP has two different expressions for \ITE: \ditet\ for constructing terms and \ditef\ for constructing formulas. \ditet\ takes a formula and two terms of the same sort as arguments. \ditef\ takes three formulas as arguments.

Since FOOL does not distinguish formulas and boolean terms, it does not require separate expressions for the formula-level and term-level \ITE. The extended version of TFF0, supported by Vampire, uses a new expression \dite, that unifies \ditet\ and \ditef. \dite\ takes a formula and two terms of the same sort as arguments. If the second and the third arguments are boolean, such \dite\  expression is equivalent to \ditef, otherwise it is equivalent to \ditet.

Consider, for example, the above definition of $\mathit{max}$. It can be encoded in the extended TFF0 as follows.
\begin{lstlisting}
tff(max, type, max: ($int * $int) > $int).
tff(max_definition, axiom,
    ![X:$int, Y:$int]:
      (max(X,Y) = $ite($greatereq(X,Y),X,Y))).
\end{lstlisting}
It uses the TPTP notation \dint\ for the sort of integers and \dgreatereq\ for the greater-than-or-equal-to comparison of two numbers.


Consider now the following valid property of $\mathit{max}$:
\begin{equation}\label{eq:ite-f-example}
  (\forall x:\Z)(\forall y:\Z)(\ite{\mathit{max}(x, y) \eql x}{x \geq y}{y \geq x}).
\end{equation}

Its encoding in the extended TFF0 can use the same \dite\ expression:
\begin{lstlisting}
![X:$int, Y:$int]: $ite(max(X,Y) = X,
                           $greatereq(X,Y),
                           $greatereq(Y,X)).
\end{lstlisting}

Note that TFF0 without \dite\ has to differentiate between terms and formulas, and so requires to use \ditet\ in~\eqref{eq:ite-t-example} and \ditef\ in~\eqref{eq:ite-f-example}.

\subsection{\LETIN}

FOOL contains \LETIN\ expressions that can be used to introduce local function definitions. They have the form
\begin{equation}\label{eq:let}
\begin{aligned}
\mathtt{let}\;&\binding{f_1(x^1_1:\sigma^1_1,\ldots,x^1_{n_1}:\sigma^1_{n_1})}{s_1};\\
              &\ldots\\
              &\binding{f_m(x^m_1:\sigma^m_1,\ldots,x^m_{n_m}:\sigma^m_{n_m})}{s_m}\\
 \mathtt{in}\;&t,
\end{aligned}
\end{equation}
where
\begin{enumerate}
  \item $m \geq 1$;
  \item $f_1,\ldots,f_m$ are pairwise distinct function symbols;
  \item $n_i \geq 0$ for each $1 \leq i \leq m$;
  \item $x^i_1\ldots,x^i_{n_i}$ are pairwise distinct variables for each $1 \leq i \leq m$; and
  \item $s_1,\ldots,s_m$ and $t$ are terms.
\end{enumerate}


The semantics of \LETIN\ expressions in FOOL mirrors the semantics of simultaneous non-recursive local definitions in programming languages. That is, $s_1,\ldots,s_m$ do not use the bindings of $f_1,\ldots,f_m$ created by this definition.

Note that an expression of the form \eqref{eq:let} is not in general equivalent to $m$ nested \LETIN s
\begin{equation}\label{eq:let-singles}
\begin{aligned}
&\mathtt{let}\;\binding{f_1(x^1_1:\sigma^1_1,\ldots,x^1_{n_1}:\sigma^1_{n_1})}{s_1}\;\mathtt{in}\\
&\quad\;\ddots\\
&\quad\quad\;\mathtt{let}\;\binding{f_m(x^m_1:\sigma^m_1,\ldots,x^m_{n_m}:\sigma^m_{n_m})}{s_m}\;\mathtt{in}\\
&\quad\quad\quad t.
\end{aligned}
\end{equation}
The main application of \LETIN\ expressions is in problems coming from program analysis, namely modelling of assignments. Consider for example the following code snippet featuring operations over an integer \verb'array'.
\begin{verbatim}
array[3] := 5;
array[2] + array[3];
\end{verbatim}
It can be translated to FOOL in the following way. We represent the integer array as an uninterpreted function $\arrayt$ of the type $\Z \to \Z$ that maps an index to the array element at that index. The assignment of an array element can be translated to a combination of \LETIN\ and \ITE.
\begin{equation}\label{eq:let-function-example}
\begin{aligned}
  &\mathtt{let}\;\binding{\arrayt(i:\Z)}{\ite{i \eql 3}{5}{\arrayt(i)}}\;\mathtt{in}\\
  &\quad\arrayt(2) + \arrayt(3)
\end{aligned}
\end{equation}


Multiple bindings in a \LETIN\ expression can be used to concisely express simultaneous assignments that otherwise would require renaming. In the following example, constants $a$ and $b$ are swapped by a \LETIN\ expression. The resulting formula is equivalent to $f(b, a)$.
\begin{equation}\label{eq:parallel-let-example}
\letinpar{a}{b}{b}{a}{f(a, b)}
\end{equation}

In order to handle \LETIN\ expressions Vampire translates them to FOL. This is done in three stages for each expression in \eqref{eq:let}.
\begin{enumerate}
  \item For each function symbol $f_i$ where $0 \leq i < m$ that occurs freely in any of $s_{i+1},\ldots,s_m$, introduce a fresh function symbol $g_i$. Replace all free occurrences of $f_i$ in $t$ by $g_i$.
  \item Replace the \LETIN\ expression by an equivalent one of the form \eqref{eq:let-singles}. This is possible because the necessary condition was satisfied by the previous step.
  \item Apply a translation to each of the \LETIN\ expression with a single binding, starting with the innermost one.
\end{enumerate}

The translation of an expression of the form $$\letin{f(x_1:\sigma_1,\ldots,x_n:\sigma_n)}{s}{t}$$ is done by the following sequence of steps. Let $y_1,\ldots,y_m$ be all free variables of $s$ and $t$, and $\tau_1,\ldots,\tau_m$ be their sorts. Note that the variables in $x_1,\ldots,x_n$ are not necessarily disjoint from the variables in $y_1,\ldots,y_m$. Let $\sigma_0$ be the sort of $s$. The steps of translation depend on whether $\sigma_0$ is $\bool$ and not. If $\sigma_0$ is not $\bool$, Vampire
\begin{enumerate}
  \item introduces a fresh function symbol $g$ of the type $$\sigma_1 \times \ldots \times \sigma_n \times \tau_1 \times \ldots \times \tau_m \to \sigma_0;$$
  \item adds to the set of assumptions the definition
  \begin{align*}
    &(\forall z_1:\sigma_1)\ldots(\forall z_n:\sigma_n) (\forall y_1:\tau_1)\ldots(\forall y_m:\tau_m)\\
    &\quad(g(z_1,\ldots,z_n,y_1,\ldots,y_m) \eql s'),
  \end{align*} where $z_1,\ldots,z_n$ is a fresh sequence of variables and $s'$ is  obtained from $s$ by replacing all free occurrences of $x_1,\ldots,x_n$ by $z_1,\ldots,z_n$, respectively; and
  \item replaces $\letin{f(x_1:\sigma_1,\ldots,x_n:\sigma_n)}{s}{t}$ by $t'$, where $t'$ is obtained from $t$ by replacing all bound occurrences of $y_1,\ldots,y_m$ by fresh variables and each application $f(t_1, \ldots, t_n)$ of a free occurrence of $f$ by $g(t_1, \ldots, t_n,\allowbreak y_1, \ldots, y_m)$.
\end{enumerate}

If $\sigma_0$ is $\bool$, the steps of translation are different:
\begin{enumerate}
  \item a fresh predicate symbol of the type \[\sigma_1 \times \ldots \times \sigma_n \times \tau_1 \times \ldots \times \tau_m\] is introduced instead;
  \item the added definition uses equivalence instead of equality.
\end{enumerate}

For example, after this translation \eqref{eq:let-function-example} becomes $g(2) + g(3)$, where $g$ is a fresh function symbol of the type $\Z \to \Z$ defined by the following formula: $$(\forall i:\Z)(g(i) \eql \ite{i \eql 3}{5}{\arrayt(i)}).$$

The example~\eqref{eq:parallel-let-example} is translated in the following way. First, the \LETIN\ expression is translated to the form~\eqref{eq:let-singles}. The constant $a$ has a free occurrence in the body of $b$, therefore it is replaced by a fresh constant $a'$. The formula \eqref{eq:parallel-let-example} becomes
\begin{equation*}
\begin{aligned}
  &\mathtt{let}\;\binding{a'}{b}\;\mathtt{in}\\
  &\quad\mathtt{let}\;\binding{b}{a}\;\mathtt{in}\\
  &\quad\quad\ f(a', b).
\end{aligned}
\end{equation*}
Then, the translation is applied to both \LETIN\ expressions with a single binding and the resulting formula becomes $f(a'', b')$, where $a''$ and $b'$ are fresh constants, defined by formulas $a'' \eql b$ and $b' \eql a$.

TPTP has four different expressions for \LETIN: \dlettt and \dletft\ for constructing terms, and \dlettf\  and \dletff\  for constructing formulas. All of them denote a single binding. \dlettt\ and \dlettf\ denote a binding of a function symbol, whereas \dletft\ and \dletff\ denote a binding of a predicate symbol. All four expressions take a (possibly universally quantified) equation as the first argument and a term (in case of \dlettt\ and \dletft) or a formula (in case of \dlettf\ and \dletff) as the second argument. TPTP does not provide any notation for \LETIN\  expressions with multiple bindings.

Similarly to \ITE, \LETIN\ expressions in FOOL do not need different notation for terms and formulas. The modification of TFF0 supported by Vampire introduces a new \dlet\ expression, that unifies \dlettt, \dletft, \dlettf\ and \dletff, and extends them to support multiple bindings. Depending on whether the binding is of a function or predicate symbol and whether the second argument of the expression is term or formula, a \dlet\ expression is equivalent to one of \dlettt, \dletft, \dlettf\ and \dletff.

The new \dlet\ expressions use different syntax for bindings. Instead of a quantified equation, they use the following syntax: a function symbol possibly followed by a list of variable arguments in parenthesis, followed by the \lstinline':=' operator and the body of the binding. Similarly to quantified variables, variable arguments are separated with commas and each variable might include a sort declaration. A sort declaration can be omitted, in which case the variable is assumed to the be of the sort of individuals (\verb|$i|).

Formula \eqref{eq:let-function-example} can be written in the extended TFF0 with the TPTP interpreted function \dsum, representing integer addition, as follows:
\begin{lstlisting}
$let(array(I:$int) := $ite(I = 3, 5, array(I)),
     $sum(array(2), array(3))).
\end{lstlisting}


The same \dlet\ expression can be used for multiple bindings. For that, the bindings should be separated by a semicolon and passed as the first argument. The formula~\eqref{eq:parallel-let-example} can be written using \dlet\ as follows.
\begin{lstlisting}
$let(a := b; b := a, f(a,b)))
\end{lstlisting}

Overall, \dite\ and \dlet\ expressions provide a more concise syntax for TPTP formulas than the TFF0 variations of \ITE\ and \LETIN\  expressions. To illustrate this point, consider the following snippet of TPTP code, taken from the TPTP problem \mbox{SYN000\_2}.
\begin{lstlisting}
tff(let_binders, axiom, ![X:$i]:
    $let_ff(![Y1:$i, Y2:$i]: (q(Y1, Y2) <=> p(Y1)),
      q($let_tt(![Z1:$i]:
          (f(Z1) = g(Z1,b)), f(a)), X) &
      p($let_ft(![Y3:$i, Y4:$i]: (q(Y3,Y4) <=>
          $ite_f(Y3 = Y4, q(a, a), q(Y3, Y4))),
          $ite_t(q(b, b), f(a), f(X)))))).
\end{lstlisting}

It uses both of the TFF0 variations of \ITE\ and three different variations of \LETIN. The same snippet can be expressed more concisely using \dite\ and \dlet\ expressions.
\begin{lstlisting}
tff(let_binders, axiom, ![X:$i]:
    $let(q(Y1,Y2) := p(Y1),
      q($let(f(Z1) := g(Z1,b), f(a)), X) &
      p($let(q(Y3,Y4) :=
               $ite(Y3 = Y4, q(a,a), q(Y3,Y4))),
          $ite(q(b,b), f(a), f(X)))))).
\end{lstlisting}

\section{Polymorphic Theory of Arrays}
\label{sect:arrays}

Using built-in arrays and reasoning in the first-order theory of arrays are common in program analysis, for example for finding loop invariants in programs using arrays~\cite{fase2009}. Previous versions of Vampire supported theories of integer arrays and arrays of integer arrays~\cite{cav13}. No other array sorts were supported and in order to implement one it would be necessary to hardcode a new sort and add the theory axioms corresponding to that sort. In this section we describe a polymorphic theory of arrays implemented in Vampire.

\subsection{Definition}
The polymorphic theory of arrays is the union of theories of arrays parametrised by two sorts: sort $\tau$ of indexes and sort $\sigma$ of values. It would have been proper to call these theories the theories of maps from $\tau$ to $\sigma$, however we decided to call them arrays for the sake of compatibility with arrays as defined in SMT-LIB.

A theory of arrays is a first-order theory that contains a sort
$\arrayt(\tau,\sigma)$, function symbols $\selectf :
\arrayt(\tau,\sigma) \times \tau \to \sigma$ and $\storef :
\arrayt(\tau,\sigma) \times \tau \times \sigma \to
\arrayt(\tau,\sigma)$, and three axioms.
The function symbol $\selectf$ represents a binary operation of
extracting an array element by its index.
The function symbol $\storef$ represents a ternary operation of updating an array at a given index with a given value. The array axioms are:
\begin{enumerate}
  \item read-over-write 1
        \begin{align*}
          &(\forall a:\arrayt(\tau,\sigma))(\forall v:\sigma)(\forall i:\tau)(\forall j:\tau)\\
          &\quad(i \eql j \implies \select{\store{a}{i}{v}}{j} \eql v);
        \end{align*}
  \item read-over-write 2
        \begin{align*}
          &(\forall a:\arrayt(\tau,\sigma))(\forall v:\sigma)(\forall i:\tau)(\forall j:\tau)\\
          &\quad(i \neql j \implies \select{\store{a}{i}{v}}{j} \eql \select{a}{j});
        \end{align*}
  \item extensionality
        \begin{align*}
          &(\forall a:\arrayt(\tau,\sigma))(\forall b:\arrayt(\tau,\sigma))\\
          &\quad((\forall i:\tau)(\select{a}{i} \eql \select{b}{i}) \implies a\eql b).
        \end{align*}
\end{enumerate}
We will call every concrete instance of the theory of arrays for
concrete sorts $\tau$ and $\sigma$ the \emph{$(\tau,\sigma)$-instance}.


One can use the polymorphic theory of arrays to express program properties. Recall the code snippet involving arrays mentioned in Section~\ref{sect:fool}:
\begin{verbatim}
array[3] := 5;
array[2] + array[3];
\end{verbatim}
Formula~\eqref{eq:let-function-example} used an interpreted function to represent the array in this code. We can alternatively use arrays to represent it as follows
\begin{equation}\label{eq:arrays-example}
\begin{aligned}
&\mathtt{let}\;\binding{\arrayt}{\store{array}{3}{5}}\;\mathtt{in}\\
&\quad\select{\arrayt}{2} + \select{\arrayt}{3}
\end{aligned}
\end{equation}

\subsection{Implementation in Vampire}

Vampire implements reasoning in the polymorphic theory of arrays by adding corresponding sorts axioms when the input uses array sorts and/or functions.

Whenever the input problem uses a sort $\arrayt(\tau,\sigma)$, Vampire adds this sort and function symbols $\selectf$ and $\storef$ of the types $\arrayt(\tau,\sigma) \times \tau \to \sigma$ and $\arrayt(\tau,\sigma) \times \tau \times \sigma \to \arrayt(\tau,\sigma)$, respectively.

If the input problem contains $\storef$, Vampire adds the following axioms for the sorts $\tau$ and $\sigma$ used in the corresponding array theory instance:

\begin{equation}\label{eq:array-axiom-1}
  \begin{aligned}
    &(\forall a:\arrayt(\tau,\sigma))(\forall i:\tau)(\forall v:\sigma)\\
    &\quad(\select{\store{a}{i}{v}}{i} \eql v)
  \end{aligned}
\end{equation}
\begin{equation}\label{eq:array-axiom-2}
  \begin{aligned}
    &(\forall a:\arrayt(\tau,\sigma))(\forall i:\tau)(\forall j:\tau)(\forall v:\sigma)\\
    &\quad(i \neql j \implies \select{\store{a}{i}{v}}{j} \eql \select{a}{j})
  \end{aligned}
\end{equation}
\begin{equation}\label{eq:array-axiom-3}
  \begin{aligned}
    &(\forall a:\arrayt(\tau,\sigma))(\forall b:\arrayt(\tau,\sigma))\\
    &\quad(a \not\eql b \implies (\exists i:\tau)(\select{a}{i} \neql \select{b}{i}))
  \end{aligned}
\end{equation}
These axioms are equivalent to the axioms read-over-write~1, read-over-write~2 and extensionality.

If the input contains only $\selectf$ but not $\storef$ for this instance, then only extensionality \eqref{eq:array-axiom-3} is added.

Theory axioms are not added when the \verb'--theory_axioms' option is set to \verb'off' (the default value is \verb'on'), which leaves an option for the user to try her or his own axiomatisation of arrays.

Vampire uses the extensionality resolution rule~\cite{ATVA14} to efficiently reason with the extensionality axiom.

To express arrays, the TPTP syntax extension supported by Vampire
allows, for every pair of sorts $\tau$ and $\sigma$, denoted by
\lstinline't' and \lstinline's' in the TFF0 syntax, to denote the sort
$\arrayt(\tau,\sigma)$ by \darray{\lstinline's'}{\lstinline't'}. Function symbols $\selectf$
and $\storef$ can be expressed as ad-hoc polymorphic \dselect\ and
\texttt{\$store}, respectively for every pairs of sorts
$\tau,\sigma$. Previously,
the theories of integer arrays and arrays of integer arrays were
represented as sorts \darrayone\ and \darraytwo\ in Vampire,
with the corresponding sort-specific function symbols \dselectone, \dselecttwo, \dstoreone\  and
\dstoretwo. Our new implementation in Vampire, with
support for the polymorphic theory of arrays, deprecates these
two concrete array theories. Instead, one can now use the sorts
\darray{\dint}{\dint} and \darray{\dint}{\darray{\dint}{\dint}}.
For example, formula~\eqref{eq:arrays-example} can be written in the extended TFF0 syntax as follows:
\begin{lstlisting}
$let(array := $store(array,3,5),
     $sum($select(array,2), $select(array,3))).
\end{lstlisting}

\subsection{Theory of Boolean Arrays}

An interesting special case of the polymorphic theory of arrays is the theory of boolean arrays. In that theory the $\selectf$ function has the type $\arrayt(\tau,\bool) \times \tau \to \bool$ and the $\storef$ function has the type $\arrayt(\tau,\bool) \times \tau \times \bool \to \arrayt(\tau,\bool)$. This means that applications of $\selectf$ can be used as formulas and $\storef$ can have a formula as the third argument.

Vampire implements the theory of booleans arrays similarly to other sorts, by adding theory axioms when the option \verb'--theory_axioms' is enabled. However, the theory axioms are different for the following reason. The axioms of the theory of boolean arrays are syntactically correct in FOOL but not in FOL, because they use quantification over booleans. However, Vampire adds theory axioms only after a translation of FOOL to FOL. For this reason, Vampire uses the following set of axioms for boolean arrays:
\begin{equation*}
  \begin{aligned}
    &(\forall a:\arrayt(\tau,\bool))(\forall i:\tau)(\forall v:\bool)\\
    &\quad(\select{\store{a}{i}{v}}{i} \liff (v \eql \true))
  \end{aligned}
\end{equation*}
\begin{equation*}
  \begin{aligned}
    &(\forall a:\arrayt(\tau,\bool))(\forall i:\tau)(\forall j:\tau)(\forall v:\bool)\\
    &\quad(i \neql j \implies \select{\store{a}{i}{v}}{j} \liff \select{a}{j})
  \end{aligned}
\end{equation*}
\begin{equation*}
  \begin{aligned}
    &(\forall a:\arrayt(\tau,\bool))(\forall b:\arrayt(\tau,\bool))\\
    &\quad(a \not\eql b \implies (\exists i:\tau)(\select{a}{i} \oplus \select{b}{i}))
  \end{aligned}
\end{equation*}
where $\oplus$ is the ``exclusive or'' connective.

\newcommand{\encrypt}{\mathit{encrypt}}
\newcommand{\key}{\mathit{key}}
\newcommand{\msg}{\mathit{message}}
\newcommand{\plaintext}{\mathit{plaintext}}
\newcommand{\cipher}{\mathit{cipher}}

One can use the theory of boolean arrays, for example, to express properties of bit vectors. In the following example we give a formalisation of a basic property of XOR encryption, where the key, the message and the cipher are bit vectors. Let $\encrypt$ be a function of the type $\arrayt(\Z,\bool) \times \arrayt(\Z,\bool) \to \arrayt(\Z,\bool)$. We will write $\encrypt(\msg, \key)$ to denote the result of bit-wise application of the XOR operation to $\msg$ and $\key$. For simplicity we will assume that the message and the key are of equal length. The definition of $\encrypt$ can be expressed with the following axiom:
\begin{align*}
  &(\forall \msg:\arrayt(\Z,\bool))(\forall \key:\arrayt(\Z,\bool))(\forall i:\Z)\\
  &\quad(\select{\encrypt(\msg,\key)}{i} \eql \\
  &\quad\quad\select{\msg}{i} \oplus \select{\key}{i}).
\end{align*}

An important property of XOR encryption is its vulnerability to the known plaintext attack. It means that knowing a message and its cipher, one can obtain the key that was used to encrypt the message by encrypting the message with the cipher. This property can be expressed by the following formula.
\begin{align*}
  &(\forall \plaintext:\arrayt(\Z,\bool))(\forall \cipher:\arrayt(\Z,\bool))\\
  &\quad(\forall \key:\arrayt(\Z,\bool))(\cipher \eql \encrypt(\plaintext,\key) \implies\\
  &\quad\quad\key \eql \encrypt(\plaintext,\cipher))
\end{align*}

The sort $\arrayt(\Z,\bool)$ is represented in the extended TFF0 syntax as \darray{\dint}{\dbool}. The presented property of XOR encryption can be expressed in the extended TFF0 in the following way.
\begin{lstlisting}
tff(encrypt, type, encrypt: ($array($int,$o) *
    $array($int,$o)) > $array($int,$o)).

tff(xor_encryption, axiom,
    ![Message:$array($int,$o),
      Key:$array($int,$o), I:$int]:
      ($select(encrypt(Message, Key), I) =
        ($select(Message, I) <~> $select(Key,I)))).

tff(known_plaintext_attack, conjecture,
    ![Plaintext:$array($int,$o),
      Cipher:$array($int,$o), Key:$array($int,$o)]:
        ((Cipher = encrypt(Plaintext, Key)) =>
          (Key = encrypt(Plaintext, Cipher)))).
\end{lstlisting}

\section{Program Analysis with the New Extensions}
\label{sect:example}

\begin{figure*}[tb]
{\small
\begin{tabular}{ll}
\begin{minipage}[t]{0.35\textwidth}
\begin{tabbing}
res \ass\ x;\\
\IF\ (x $>$ y) \\\inc
\THEN\ max \ass\ x;\\
\ELSE\ max \ass\ y;\\\dec
\IF\ (max $>$ 0) \\\inc
\THEN\ res \ass\ res $+$ max;\\
\ELSE\ res \ass\ res $-$ max;\\[.5em]\dec
\reserved{assert} res $\geq$ x
\end{tabbing}\vspace*{-.75em}
\caption{Sequence of conditionals.\label{fig:seqITE}}
\begin{tabbing}
\IF\ (x $>$ y) \\\inc
\THEN\ t \ass\ x; x \ass\ y; y \ass\ t;\\\dec
\reserved{assert} y $\geq$ x
\end{tabbing}\vspace*{-.75em}
\caption{Updating multiple variables.\label{fig:tmpSwap}}
\end{minipage}
&
\begin{minipage}[t]{0.6\textwidth}
\begin{tabbing}
$a$ \ass\ $0$; $b$ \ass\ $0$; $c$ \ass\ $0$; \\[.5em]
\reserved{invariant} a = b + c $\wedge$ \\
{\color{white}\reserved{invariant}} a $\geq$ 0 $\wedge$ b $\geq$ 0 $\wedge$ c $\geq$
0 $\wedge$ a $\leq$ k $\wedge$ \\
{\color{white}\reserved{invariant}} $(\forall p) (0\leq p<b \implies
(\exists i) (0 \leq i < a \wedge A[i] > 0 \wedge B[p] = A[i]))$\\[1em]
\WHILE\ ($a \leq k$) \DO \\ \inc
  \IF\ ($A[a] > 0$) \\ \inc
    \THEN\ \=\+ $B[b]$ \ass\ $A[a]$\semicol $b$ \ass\ $b+1$\semicol \\ \dec
    \ELSE\ \=\+ $C[c]$ \ass\ $A[a]$\semicol $c$ \ass\ $c+1$\semicol \\ \dec \dec
  $a$ \ass\ $a+1$\semicol \\ \dec
\OD\\[.5em]
\reserved{assert} $(\forall p)(0 \leq p<b \implies B[p]> 0)$
\end{tabbing}\vspace*{-.75em}
\caption{Array partition.\label{fig:partition}}
\end{minipage}
\end{tabular}
}
\end{figure*}

\begin{figure*}[t]
{\small
\begin{lstlisting}
tff(x, type, x: $int).
tff(y, type, y: $int).
tff(max, type, max: $int).
tff(res, type, res: $int).
tff(res1, type, res1: $int).

tff(transition_relation, hypothesis,
    res1 = $let(res := x,
             $let(max := $ite($greater(x,y), $let(max := x, max), $let(max := y, max)),
               $let(res := $ite($greater(max,0), $let(res := $sum(res,max), res), $let(res := $difference(res,max),res)),
                 res)))).

tff(safety_property, conjecture, $greatereq(res1,x)).
\end{lstlisting}
\caption{Representation of the partial correctness statement of
  Figure~\ref{fig:seqITE}\label{fig:VampireITE}.}
}
\end{figure*}

In this section we illustrate how FOOL makes first-order theorem
provers better suited to applications in program analysis and
verification.
Firstly,  we give concrete examples exemplifying the use of FOOL for
expressing program properties. We avoid various
program analysis steps, such as SSA form computations and renaming
program variables; instead we show how program properties can directly
be expressed in FOOL.
 We also present a technique for
automatically generating the next state relation of any program with
assignments, \ITE, and sequential composition.
For doing so,  we introduce a simple extension of FOOL,
allowing for a general translation that is linear in the size of the
program.
This is a new result intended to understand which extensions of
first-order logic are adequate for naturally representing fragments of
imperative programs.

\begin{figure*}[t]
{\small
\begin{lstlisting}
tff(a, type, a: $int).
tff(b, type, b: $int).
tff(c, type, c: $int).
tff(k, type, k: $int).
tff(arrayA, type, arrayA: $array($int,$int)).
tff(arrayB, type, arrayB: $array($int,$int)).
tff(arrayC, type, arrayC: $array($int,$int)).

tff(invariant_property, hypothesis,
    inv <=> ((a = $sum(b, c)) & $greatereq(a,0) & $greatereq(b,0) & $greatereq(c,0) & $lesseq(a,k) &
             ![P:$int]: ($lesseq(0,P) & $less(P,b) =>
               (?[I:$int]: ($lesseq(0,I) & $less(I,a) &
                             $greater($select(arrayA,I),0) & $select(arrayB,P) = $select(arrayA,I)))))).

tff(safety_property, conjecture,
    (inv & ~$lesseq(a,k)) => (![P:$int]: ($lesseq(0,P) & $less(P,b) => $greater($select(arrayB,P),0)))).
\end{lstlisting}
\caption{Representation of the partial correctness statement of
  Figure~\ref{fig:partition} in Vampire\label{fig:loop_safety_Vampire}.}
}
\end{figure*}

\subsection{Encoding the next state relation}\label{sec:foolp}

Consider the program given in
Figure~\ref{fig:seqITE}, written in a C-like syntax, using a sequence
of two conditional statements.
The program first computes the maximal value $max$ of two integers $x$ and
$y$ and then adds the absolute value of $max$ to $x$. A safety assertion,
in FOL, is specified at the end of the loop, using the
{\bf assert} construct. This program is clearly safe, the
assertion is satisfied. To prove program safety, one needs to reason
about the program's transition relation, in particular reason about
conditional statements, and express the final value of
the program variable $res$. The partial correctness of the program
of Figure~\ref{fig:seqITE} can be \emph{automatically} expressed in FOOL,
and then Vampire can be used to prove program safety.
This requires us to encode (i)
the next state value of $res$ (and $max$) as a hypothesis
in the extended TFF0 syntax of FOOL,
by using the \ITE\ ({\tt \$ite}) and \LETIN\ ({\tt \$let})
constructs, and (ii)
the safety property as the conjecture to be proven by Vampire.

Figure~\ref{fig:VampireITE} shows this extended TFF0 encoding.
The use of \ITE\ and \LETIN\ constructs allows us to have a
direct  encoding of the  transition relation of
Figure~\ref{fig:seqITE} in FOOL. Note that each expression from the program appears only once in the encoding.

We now explain how the encoding of the next state values of program
variables can be generated automatically.
We consider programs using assignments \texttt{:=},
\ITE\ and sequential composition $;$.
We begin by making an assumption about the structure of programs (which we relax later). A program $P$ is in \emph{restricted form} if for any subprogram of the form \IF\ e \THEN\ $P_1$ \ELSE\ $P_2$ the subprograms $P_1$ and $P_2$ only make assignments to the same single variable. Given a program $P$ in restricted form let us define its translation $[P]$ inductively as follows:
\begin{itemize}
	\item $[$x\ASS e$]$ is $\letin{x}{e}{x}$;
	\item $[$\IF\ e \THEN\ $P_1$ \ELSE\ $P_2]$, where $P_1$ and
          $P_2$ update $x$,  is $\letin{x}{\ite{e}{[P_1]}{[P_2]}}{x}$;
	\item $[P_1$;$P_2]$ is $\mathtt{let}~D~\mathtt{in}~[P_2]$ where $[P_1]$ is $\mathtt{let}~D~\mathtt{in}~x$.
\end{itemize}
Given a program $P$, the next state value for variable $x$ can be
given by $[P$; x\ASS x$]$,
i.e. by ensuring the final statement of the program updates the
variable of interest.
The restricted form is required as conditionals must be viewed
as assignments in the translation and assignments can only be made to single variables.

To demonstrate the limitations of this restriction let us consider the simple program in Figure~\ref{fig:tmpSwap} that ensures that x is not larger than y. We cannot apply the translation as the conditional updates three variables. To generalise the approach we can extend FOOL with \emph{tuple expressions}, let us call this extension \foolp. In this extended logic the next state values for Figure~\ref{fig:tmpSwap} can be encoded as follows:
\[
\begin{array}{ll}
\mathtt{let}\; (x,y,t) =& \mathtt{if}\; x > y \;\mathtt{then} \\
&\quad\mathtt{let}\; (x,y,t) = (x,y,x) \;\mathtt{in}\\
&\quad\quad\mathtt{let}\; (x,y,t) = (y,y,t) \;\mathtt{in}\\
&\quad\quad\quad\mathtt{let}\;(x,y,t) = (x,t,t)  \;\mathtt{in}\; (x,y,t) \\
&\mathtt{else}\; (x,y,t)\\
\mathtt{in}\; (x,y,t) \\
\end{array}
\]
We now give a brief sketch of the extended logic \foolp\ and the associated translation. We omit details since its full definition and semantics would require essentially repeating definitions from \cite{FOOL}.  \foolp\ extends \fool\ by tuples; for all expressions $t_i$ of type $\sigma_i$ we can use a \emph{tuple expression} $(t_1,\ldots,t_n)$ of type $(\sigma_1,\ldots,\sigma_n)$. The logic should also include a suitable tuple projection function, which we do not discuss here.

This extension allows for a more general translation in two senses:
first, the previous restricted form is lifted; and second, it now
gives the next state values of {\it all} variables updated by the program. Given a program $P$ its translation $[P]$ will have the form $\letin{(x_1,\ldots,x_n)}{E}{(x_1,\ldots,x_n)}$, where $x_1,\ldots,x_n$ are all variables updated by $P$, that is, all variables used in the left-hand-side of an assignment. We inductively define $[P]$ as follows:
\begin{itemize}
	\item $[$x$_i$ \ASS e$]$ is $\letin{(\ldots,x_i,\ldots)}{(\ldots,e,\ldots)}{(x_1,\ldots,x_n)}$,
	\item $[$\IF\ e \THEN\ $P_1$ \ELSE\ $P_2]$ is $\mathtt{let}~(x_1,\ldots,x_n) = \mathtt{if}~e~\mathtt{then}$ $[P_1]~\mathtt{else}~[P_2]~\mathtt{in}~(x_1,\ldots,x_n)$,
		\item $[P_1$;$P_2]$ is $\mathtt{let}~D~\mathtt{in}~[P_2]$ where $[P_1]$ is $\mathtt{let}~D~\mathtt{in}~(x_1,\ldots,x_n)$.
\end{itemize}
This translation is bounded by $O(v\cdot n)$, where $v$ is the number
of variables in the program and $n$ is the program size (number of
statements) as each program statement is used once with one or two
instances of $(x_1,\ldots,x_n)$.
This becomes $O(n)$ if we assume that the number of
variables is fixed. The translation could be refined so that some introduced  \LETIN\
expressions only use a subset of program variables.
Finally, this translation preserves the semantics
of the program.

\begin{theorem}\rm
  Let $P$ be a program with variables $(x_1,\ldots,x_n)$ and let $u_1,\ldots,u_n, v_1, \ldots, v_n$ be values (where $u_i$ and $v_i$ are of the same type as $x_i$). If $P$ changes the state $\{x_1\to u_1,\ldots,x_n\to u_n\}$ to $\{x_1\to v_1,\ldots,x_n\to v_n\}$ then the value of $[P]$ in $\{x_1\to u_1,\ldots,x_n\to u_n\}$ is $(v_1,\ldots,v_n)$.
\end{theorem}

This translation encodes the next state values of program variables by
directly following the structure of the program. This leads to a
succinct representation that, importantly, does not lose any
information or attempt to translate the program too early. This allows
the theorem prover to apply its own translation to FOL that it can
handle efficiently.   While \foolp{} is not yet fully supported in
Vampire, we believe experimenting with \foolp{} on
examples coming from program analysis and verification is an
interesting task for future work.

\subsection{A program with a loop and arrays}

Let us now show the use of FOOL in Vampire for reasoning about
programs with loops. Consider the program given in
Figure~\ref{fig:partition}, written in a C-like syntax.  The program
fills an integer-valued array $B$ by the strictly positive values
of a source array $A$, and an integer-valued array $C$ with
the non-positive values of $A$. A safety assertion, in FOL, is
specified at the end of the loop, using the {\bf assert}
construct. The program of Figure~\ref{fig:partition} is clearly safe
as the assertion is satisfied when the loop is exited.
However, to prove program safety we need additional
loop properties, that is loop invariants, that hold at any loop
iteration. These can be automatically generated using existing approaches, for
example the symbol elimination method for invariant generation in
Vampire~\cite{fase2009}. In this case we use the FOL property
specified in the {\bf
  invariant} construct of  Figure~\ref{fig:partition}. This invariant
property states that at any loop iteration, (i) the  sum of visited
array elements in $A$ is the sum of visited elements in $B$ and $C$
(that is, $a = b + c$), (ii) the number of visited array
elements in $A$, $B$, $C$ is positive (that is, $a\geq 0$, $b\geq 0$,
and $c\geq 0$), with $a\leq k$, and (iii) each array element
$B[0],\ldots,B[b-1]$ is a strictly positive element in
$A$. Formulating the latter property requires quantifier alternation
in FOL, resulting in the quantified property with $\forall\exists$
listed in the invariant of  Figure~\ref{fig:partition}.
We can verify the safety of the program using Hoare-style reasoning in Vampire.
The partial correctness property is that the invariant and the negation of the loop condition implies the safety assertion.
This is the conjecture to be proven by Vampire.
Figure~\ref{fig:loop_safety_Vampire} shows the encoding in the
extended TFF0 syntax of this partial
correctness statement; note that this uses the built-in theory of
polymorphic arrays in Vampire, where $arrayA$, $arrayB$ and $arrayC$
correspond respectively to the arrays $A$, $B$ and $C$. 
%

So far,  we assumed that the given invariant in
Figure~\ref{fig:partition} is
indeed an invariant. Using \foolp{} described in
Section~\ref{sec:foolp}, we can verify the inductiveness
property of the invariant, as follows: (i) express the the transition
relation of the loop in \foolp, and (ii) prove that, if the invariant
holds at an arbitrary loop iteration $i$, then it also holds at loop
iteration $i+1$. For proving this, we can again use \foolp\ to
formulate
the next state values of loop variables in the invariant at loop
iteration $i+1$.
Moreover, \foolp{} can also be used to express formulas as
inputs to the symbol elimination method for invariant generation in
Vampire. We leave the task of using \foolp{} for invariant generation
as further work.

%

\section{Experimental Results}
\label{sect:experiments}

The extension of Vampire to support FOOL and the polymorphic theory of
arrays comprises about 3,100 lines of C++ code, of which the
translation of FOOL to FOL and FOOL paramodulation takes about 2,000
lines, changes in the parser about 500 lines and
the implementation of the polymorphic theory of arrays about 600 lines.
Our implementation is available at \url{www.cse.chalmers.se/~evgenyk/fool-experiments/} and will be included
in the forthcoming official release of Vampire.

In the sequel, by Vampire we mean its version including support for
FOOL and the polymorphic theory of arrays. We write \nofoolVampire\ for
its version with FOOL paramodulation turned off.

In this section we present experimental results obtained by running Vampire on FOOL problems. Unfortunately, no large collections of such problems are available, because FOOL was not so far supported by any first-order theorem prover. What we did was to extract such benchmarks from other collections.

\begin{enumerate}
\item We noted that many problems in the higher-order part of the TPTP library~\cite{TPTP} are FOOL problems, containing no real higher-order features. We converted them to FOOL problems.

\item We used a collection of first-order problems about (co)al\-ge\-braic datatypes, generated by the Isabelle theorem prover~\cite{Isabelle}, see Subsection~\ref{subsec:Isabelle} for more details.
\end{enumerate}
Our results are summarised in Tables~\ref{table:thf-results}--\ref{table:smt-lib-nontrivial} and discussed below. These results were obtained on a MacBook Pro with a 2,9 GHz Intel Core i5 and 8 Gb RAM, and using the time limit of 60 seconds per problem. Both the benchmarks and the results are available at \url{www.cse.chalmers.se/~evgenyk/fool-experiments/}.

\subsection{Experiments with TPTP Problems}
The higher-order part of the TPTP library contains 3036 problems. Among these problems, 134 contain either boolean arguments in function applications or quantification over booleans, but contain no lambda abstraction, higher-order sorts or higher-order equality. We used these 134 problems, since they belong to FOOL but not to FOL. We translated these problems from THF0 to the modification of TFF0, supported by Vampire using the following syntactic transformation: \begin{enumerate*}[label=(\alph*)]
\item every occurrence of the keyword \verb'thf' was replaced by \verb'tff';
\item every occurrence of a sort definition of the form \verb's_1 >  ... > s_n > s' was replaced by \verb's_1 * ... * s_n > s';
\item every occurrence of a function application of the form \verb'f @  t_1 @ ... @ t_n' was replaced by \verb'f(t_1, ..., t_n)'.
\end{enumerate*}

Out of 134 problems, 123 were marked as Theorem and 5 as
Unsatisfiable, 5 as CounterSatisfiable, and 1 as Satisfiable, using
the SZS status of TPTP. Essentially, this means that among their
satisfiability-checking analogues, 128 are unsatisfiable and 6 are
satisfiable. Vampire was run with the \verb'--mode casc' option for
unsatisfiable (Theorem and Unsatisfiable) problems and with \verb'--mode casc_sat' for satisfiable (CounterSatisfiable and Satisfiable) problems. These options correspond to the CASC competition modes of
Vampire for respectively proving validity (i.e. unsatisfiability) and
satisfiability of an input problem.

For this experiment, we compared the performance of Vampire with those of the higher-order theorem provers used in the the latest edition of CASC \cite{CASC25}:
Satallax~\cite{Satallax}, Leo-II~\cite{LeoII}, and Isabelle~\cite{Isabelle}. We note that all of them used the first-order theorem prover E~\cite{E13} for first-order reasoning (Isabelle also used several other provers).

\begin{table}
  \centering
  \begin{tabular}{lrr}
    \hline Prover & Solved & Total time on solved problems \\ \hline
    Vampire & 134 & 3.59 \\
    \nofoolVampire & 134 & 7.28 \\
    Satallax & 134 & 23.93 \\
    Leo-II & 127 & 27.42 \\
    Isabelle & 128 & 893.80
  \end{tabular}
  \caption{Runtimes in seconds of provers on the set of 134 higher-order TPTP problems.}
  \label{table:thf-results}
\end{table}

Table~\ref{table:thf-results} summarises our results on these problems. Only Vampire, \nofoolVampire\ and Satallax were able to solve all of them, while
Vampire was the fastest among all provers. We believe these results
are significant for two reasons. First, for solving these problems
previously one 
needed higher-order theorem provers , but now can they be proven using first-order reasoners. Moreover, even on such simple problems there is a clear gain from using FOOL paramodulation.


\subsection{Experiments with Algebraic Datatypes Problems}
\label{subsec:Isabelle}

For this experiment, we used 152 problems generated by the Isabelle theorem prover. These
problems express various properties of (co)algebraic datatypes and are written in the SMT-LIB~2 syntax~\cite{SMT-LIB}. All 152 problems contain quantification over booleans, boolean arguments in function/predicate applications and \ITE\ expressions. These examples were generated and given to us by Jasmin Blanchette, following the recent work on reasoning about (co)datatypes~\cite{Blanchette15}. To run the benchmark we first translated the SMT-LIB files to the TPTP syntax using the SMTtoTPTP translator~\cite{SMTLIB2TPTP} version 0.9.2.
Let us note that this version of SMTtoTPTP does not fully support the
boolean type in SMT-LIB. However, by setting the option
\verb'--keepBool' in SMTtoTPTP, we managed to translate these 152
problems into an extension of TFF0, which Vampire can read.
We also modified the source code of  SMTtoTPTP so that  \ITE\
expressions in the SMT-LIB files are not expanded but translated to \dite\
in FOOL. A similar modification would have been needed for translating
\LETIN\ expressions; however, none of our 152 examples used \LETIN.

After translating these 152 problems into an extended TFF0 syntax
supporting FOOL, we ran Vampire twice on each benchmark: once using the
option \verb'--mode casc', and once using
\verb'--mode casc_sat'.  For each problem, we recorded the
fastest successful run of Vampire. We used a similar setting for
evaluating \nofoolVampire.
In this experiment, we then compared Vampire with
the best available SMT solvers, namely with CVC4~\cite{CVC4} and
Z3~\cite{Z3}.

\begin{table}
  \centering
  \begin{tabular}{lrr}
    \hline Prover & Solved & Total time on solved problems \\ \hline
    Vampire & 59 & 26.580 \\
    Z3 & 57 & 4.291 \\
    \nofoolVampire & 56 & 26.095 \\
    CVC4 & 53 & 25.480
  \end{tabular}
  \caption{Runtimes in seconds of provers on the set of 152 algebraic datatypes problems.}
  \label{table:smt-lib-results}
\end{table}

\begin{figure}
  \vspace{-0.3em}
  \centering
  \begin{tikzpicture}
    \draw (0,0) circle (1.5cm);
    \draw (50:1cm) circle (1.55cm);
    \draw (0cm:0.8cm) circle (1.4cm);
    \node at (0.8cm:0.5cm) {$51$}; %
    \node at (-2.2cm:1.2cm) {$1$};
    \node at (1cm:-1.1cm) {$2$};
    \node at (-2cm:-1.1cm) {$3$};
    \node at (-3.8cm:-1.9cm) {$4$}; %
    \node at (-0.95cm:1.775cm) {$0$}; %
    \node at (0.45cm:1.775cm) {$1$}; %
    \node at (2.7cm:2.65cm) {Vampire};
    \node at (-3.7cm:1.8cm) {Z3};
    \node at (-1.6cm:2.3cm) {CVC4};
  \end{tikzpicture}
  \vspace{-0.3em}
  \caption{Venn diagram of the subsets of the algebraic datatypes problems, solved by Vampire, CVC4 and Z3.}
  \label{fig:smt-lib-diagram}
\end{figure}
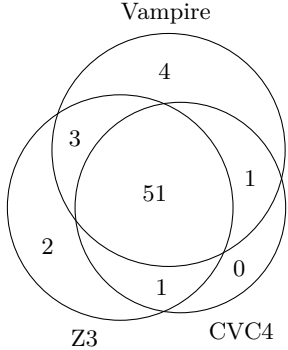

Table~\ref{table:smt-lib-results} summarises the results of our experiments on these 152 problems. Vampire solved the largest number of problems, and all problems solved by \nofoolVampire\ were also solved by Vampire.
Figure~\ref{fig:smt-lib-diagram} shows the Venn diagram of the sets of
problems solved by Vampire, CVC4 and Z3, where the numbers denote the numbers of solved problems.
All problems apart from 11 were either solved by all systems or not solved by all systems. Table~\ref{table:smt-lib-nontrivial} details performance results on these 11 problems.

Based on our experimental results shown in Tables~\ref{table:smt-lib-results} and \ref{table:smt-lib-nontrivial}, we make the following observations. On the given set of problems the implementation of FOOL reasoning in Vampire was efficient enough to compete with state-of-the-art SMT solvers. This is significant because the problems were tailored for SMT reasoning. Vampire not only solved the largest number of problems, but also yielded runtime results that are comparable with those of CVC4. Whenever successful, Z3 turned out to be faster than Vampire; we believe this is because of the sophisticated preprocessing steps in Z3. Improving FOOL preprocessing in Vampire, for example for more efficient CNF translation of FOOL formulas, is an interesting task for further research. We note that the usage of FOOL paramodulation showed improvement on the number of solved problems.

\newcommand{\timeout}{---}
\newcommand{\gaveup}{---}
\begin{table*}[t]
  \centering
  \begin{tabular}{lrrr}
    \hline Problem & Vampire & CVC4 & Z3 \\ \hline
    \verb'afp/abstract_completeness/1830522' & \timeout & \timeout & 0.172 \\
    \verb'afp/bindag/2193162' & \timeout & \gaveup & 0.388 \\
    \verb'afp/coinductive_stream/2123602' & \timeout & 0.373 & 0.101 \\
    \verb'afp/coinductive_stream/2418361' & 3.392 & \timeout & \timeout \\
    \verb'afp/huffman/1811490' & 0.023 & \gaveup & \timeout \\
    \verb'afp/huffman/1894268' & 0.025 & \gaveup & 0.052 \\
    \verb'distro/gram_lang/3158791' & 0.047 & 0.179 & \timeout \\
    \verb'distro/koenig/1759255' & 0.070 & \timeout & \timeout \\
    \verb'distro/rbt_impl/1721121' & 4.523 & \timeout & \timeout \\
    \verb'distro/rbt_impl/2522528' & 0.853 & \gaveup & 0.064 \\
    \verb'gandl/bird_bnf/1920088'	& 0.037 & \timeout & 0.077
  \end{tabular}
  \caption{Runtimes in seconds of provers on selected algebraic datatypes problems. Dashes mean the solver failed to find a solution.}
  \label{table:smt-lib-nontrivial}
\end{table*}

\section{Related Work}
\label{sect:related}

FOOL was introduced in our previous work~\cite{FOOL}. This also presented a translation from FOOL to the ordinary first-order logic, and FOOL paramodulation. In this paper we describe the first practical implementation of FOOL and FOOL paramodulation.

Superposition theorem proving in finite domains, such as the boolean domain, is also discussed in~\cite{HillenbrandWeidenbach13}. The approach of~\cite{HillenbrandWeidenbach13} sometimes falls back to enumerating instances of a clause by instantiating finite domain variables with all elements of the corresponding domains. Nevertheless, it allows one to also handle finite domains with more than two elements. One can also generalise our approach to arbitrary finite domains by using binary encodings of finite domains. However, this will necessarily result in loss of efficiency, since a single variable over a domain with $2^k$ elements will become $k$ variables in our approach, and similarly for function arguments.
Although \cite{HillenbrandWeidenbach13} reports preliminary results with the theorem prover SPASS, we could not make an experimental comparison since the SPASS implementation has not yet been made public.

Handling boolean terms as formulas is common in the SMT community. The SMT-LIB project~\cite{SMT-LIB} defines its core logic as first-order logic extended with the distinguished first-class boolean sort and the \LETIN\ expression used for local bindings of variables. The language of FOOL extends the SMT-LIB core language with local function definitions, using \LETIN\ expressions defining functions of arbitrary, and not just zero, arity.

A recent work \cite{SMTLIB2TPTP} presents SMTtoTPTP, a translator from SMT-LIB to TPTP. SMTtoTPTP does not fully support boolean sort, however one can use SMTtoTPTP with the \verb'--keepBool' option to translate SMT-LIB problems to the extended TFF0 syntax, supported by Vampire.

Our implementation of the polymorphic theory of arrays uses a syntax that coincides with the TPTP's own syntax for polymorphically typed first-order logic TFF1~\cite{tff1}.

\section{Conclusion and Future Work}
\label{sect:future}

We presented new features recently implemented in Vampire. They include FOOL: the extension of first-order logic by a first-class boolean sort, \ITE\ and \LETIN\ expressions, and polymorphic arrays. Vampire implements FOOL by translating FOOL formulas into FOL formulas. We described how this translation is done for each of the new features. Furthermore, we described a modification of the superposition calculus by FOOL paramodulation that makes Vampire reasoning in FOOL more efficient. 
We also give a simple extension to FOOL, 
allowing to express the next state relation of a program as a boolean formula which is linear in the size of the program.

Neither FOOL nor polymorphic arrays can be expressed in TFF0. In order to support them Vampire uses a modification of the TFF0 syntax with the following features:
\begin{enumerate}
  \item the boolean sort \tptpo\ can be used as the sort of arguments and quantifiers;
  \item boolean variables can be used as formulas, and formulas can be used as boolean arguments;
  \item \ITE\ expressions are represented using a single keyword \dite\ rather than two different keywords \ditet\ and \ditef;
  \item \LETIN\ expressions are represented using a single keyword \dlet\ rather than four different keywords \dlettt, \dlettf, \dletft\ and \dletff;
  \item \darraySymb, \dselect\ and \dstore\ are used to represent arrays of arbitrary types.
\end{enumerate}
Our experimental results have shown that our implementation, and especially FOOL paramodulation, are efficient and can be used to solve hard problems.

Many program analysis problems, problems used in the SMT community, and problems generated by interactive provers, which previously required (sometimes complex) ad hoc translations to first-order logic, can now be understood by Vampire without any translation. Furthermore, Vampire can be used to translate them to the standard TPTP without \ITE\ and \LETIN\ expressions, that is, the format understood by essentially all modern first-order theorem provers and used at recent CASC competitions. One should simply use \texttt{--mode preprocess} and Vampire will output the translated problem to \texttt{stdout} in the TPTP syntax. 

The translation to FOL described here is only the first step to the efficient handling of FOOL. It can be considerably improved. For example, the translation of \LETIN\ expressions always introduces a fresh function symbol together with a definition for it, whereas in some cases inlining the function would produce smaller clauses. Development of a better translation of FOOL is an important future work.

FOOL can be regarded as the smallest superset of the SMT-LIB~2 Core language and TFF0. A native implementation of an SMT-LIB parser in Vampire is an interesting future work. Note that such an implementation can also be used to translate SMT-LIB to FOOL or to FOL.

Another interesting future work is extending FOOL to handle polymorphism and implementing it in Vampire. This would allow us to parse and prove problems expressed in the TFF1~\cite{tff1} syntax. Note that the current usage of \darraySymb\ conforms with the TFF1 syntax for type constructors.

\acks
We acknowledge funding from the Austrian FWF National Research Network RiSE
S11409-N23, the Swedish VR grant D049770 - GenPro, the
Wallenberg Academy Fellowship 2014, and the EPSRC grant ``Reasoning in Verification and Security''.

%
\label{sect:bib}
\bibliographystyle{abbrvnat}
\bibliography{refs}


\end{document}